\documentclass[a4paper,11pt]{article}
\pdfoutput=1 
\usepackage{leftidx}
\usepackage{jheppub} 
\usepackage[T1]{fontenc} 
\usepackage{xcolor}
\usepackage{subfigure}
\usepackage[export]{adjustbox}
\usepackage{amsmath}

\newcommand{\ket}[1]{\left|#1\right\rangle}
\newcommand{\bra}[1]{\left\langle #1\right|}

\newcommand{\be}{\begin{equation}}
\newcommand{\ee}{\end{equation}}

\def\text#1{{\rm #1}}

\def\bra#1{\left\langle#1\right|}
\def\ket#1{\left|#1\right\rangle}
\def\braket#1#2{\left\langle #1\middle|#2\right\rangle}

\def\abs#1{\left|#1\right|}
\def\kc#1{\left(#1\right)}
\def\kd#1{\left[#1\right]}
\def\ke#1{\left\{#1\right\}}

\def\be{\begin{equation}}       \def\ee{\end{equation}}
\def\bea{\begin{eqnarray}}      \def\eea{\end{eqnarray}}
\def\ba{\begin{array} }
\def\ea{\end{array} }

\def\nn{\nonumber}
\def\pa{\partial}
\def\=>{\Rightarrow}
\def\>{\rightarrow}

\def\sket#1{\left|#1\right)}
\def\sbra#1{\left(#1\right|}

\newcounter{facts}[section]

\renewcommand{\thefacts}{\arabic{facts}}

\title{\boldmath Emergent bulk gauge field in random tensor networks}

\author{Xiao-Liang Qi}

\affiliation{Stanford Institute for Theoretical Physics, Stanford University, Stanford, California 94305, USA}

\abstract{Random tensor network states are toy models for holographic duality, which have entanglement properties determined by graph geometry. In this paper, we propose a generalization of the random tensor network states which describe an ensemble of states preserving a given global symmetry. We show that Renyi entropy for this family of states can be described by a quantum extremal surface formula, with corrections to the area law term determined by a bulk gauge theory wavefunction. This provides a toy model of the correspondence between boundary global symmetry and bulk gauge symmetry in holographic duality. We discuss the boundary physical consequences of the bulk deconfined and confined phases. }

\begin{document} 
\maketitle

\section{Introduction}

One of the groundbreaking works of Prof. Chen-Ning Yang is the Yang-Mills theory of non-Abelian gauge fields\cite{yang1954conservation}. Except for gravity, all fundamental interactions in our world are mediated by gauge fields. In 1997, Juan Maldaceana proposed that the $N=4$ supersymmetric Yang-Mills theory in $4$ spacetime dimensions is dual to a $5$-dimensional supergravity theory with asymptotically anti-de Sitter (AdS) spacetime\cite{maldacena1999large,gubser1998gauge,witten1998anti}. This is known as AdS/CFT (conformal field theory) correspondence, or holographic duality, which has been supported by more and more evidences, and lead to a lot of new progress in our understanding to quantum gravity. The holographic duality suggests that gauge invariance principle not only describes other fundamental forces but also plays an essential role in understanding gravity. This paper is a toy model study of a particular aspect of holographic duality---the correspondence between boundary global symmetry and bulk gauge symmetry.

Quantum information concepts plays an essential role in understanding holographic duality. The Ryu-Takayanagi (RT) formula\cite{ryu2006holographic} pointed out that von Neumann entropy of the boundary theory is dual to the extremal surface area in Planck unit in the bulk theory. 
To understand more explicitly this relation between geometry and entanglement, tensor network states have been proposed as toy models of holographic duality\cite{swingle2012entanglement,qi2013exact,pastawski2015holographic,yang2016bidirectional,hayden2016holographic}. Tensor network is a method widely used in condensed matter physics, which is a general construction of many-body states from few body building blocks.\cite{cirac2021matrix} A general tensor network state is obtained by first preparing maximally entangled EPR pairs along each edge of a graph, and then apply a vertex projection onto the qubits at that vertex. The result of this procedure is a quantum many-body state of remaining qubits at the end of dangling edges of the graph. Ref. \cite{pastawski2015holographic} shows that tensor networks with a hyperbolic graph geometry and particular types of ``perfect tensors" satisfy Ryu-Takayanagi formula, and also correctly reproduced the ``quantum error correction" nature of the bulk-boundary correspondence pointed out in Ref. \cite{almheiri2015bulk}. Ref. \cite{hayden2016holographic} shows that large bond-dimension random tensor networks (RTN), with vertex projections defined by random states, approximately satisfy the Ryu-Takayanagi formula and its quantum corrections, for general graph geometries. Renyi entropy calculation of random tensor networks can be mapped to partition functions of classical statistical models, which allowed us to understand various properties of RTN systematically on general graphs. 

One problem with RTN is that, due to the randomness, they do not preserve any symmetry of the boundary. In holographic duality, the proposal is that a global symmetry in the boundary theory corresponds to a gauge symmetry in the bulk\cite{witten1998anti}, while energy conservation of the boundary (if the boundary is time translation invariant) is related to bulk diffeomorphism. Therefore a natural question for making more realistic toy models of holographic duality is to look for generalizations of tensor networks that preserve symmetries on the boundary, and see if there is an analog of the global-gauge correspondence. 

In this paper, we generalize the RTN model to random tensors preserving a given global symmetry. Our construction leads to boundary states with global symmetry, and the bulk theory contains an emergent gauge field. We will discuss physical consequences of this bulk gauge theory. In particular, the bulk gauge theory could be in a topological phase with nontrivial topological entropy\cite{kitaev2006topological,levin2006detecting}. We will discuss the boundary interpretation of the topological entropy. 

The remainder of this paper is organized as follows. Sec. \ref{sec:RTN} defines the RTN with global symmetry, and discuss its relation with bulk gauge field. Sec. \ref{sec:gauge} provides further background information about lattice gauge theory and Levin-Wen model\cite{levin2005string}, and discuss the topological entropy in the bulk and its boundary interpretation. Finally, Sec. \ref{sec:conclusion} contains the conclusion and further discussions.  

\section{Random tensor network states with global symmetry}
\label{sec:RTN}

\subsection{Definition of the states}
A tensor network is defined by a pair of quantum states
\begin{align}
    \ket{\Psi_P}&\in\mathbb{H}_I\otimes\mathbb{H}_B,~\ket{V}\in\mathbb{H}_I\\
    \ket{\Psi}&=\braket{V}{\Psi_P}
\end{align}
with $\mathbb{H}_I$ the bulk (interior) Hilbert space, and $\mathbb{H}_B$ the boundary Hilbert space. For standard tensor networks, one first choose a graph with vertices $x$ and edges $\overline{xy}$. The set of $x$ is divided into the interior $I$ and the boundary $B$.  $\ket{\Psi_P}=\otimes_{xy}\ket{xy}$ is defined as a product of EPR pairs, one on each link of a graph. $\ket{V}=\otimes_{x\in I}\ket{V_x}$ is a product of vertex states at each bulk vertex. Here $\ket{V_x}$ is defined in a Hilbert space of site $x$, which is a product of all qubits at that site, one for each edge connected with $x$. If we express both $\ket{V_x}$ and $\ket{xy}$ in a chosen basis, the wavefunction of $\ket{V_x}$ are vertex tensors, while the state $\ket{xy}$ gives a ``metric" for contracting indices of tensors on neighboring sites. The wavefunction of the boundary sites is given by contracting the tensors on all internal indices. In RTN\cite{hayden2016holographic} $\ket{\Psi_P}$ can also be generalized to include non-EPR pair states.  

In Ref. \cite{hayden2016holographic}, the vertex states $\ket{V_x}=U_x\ket{0_x}$ are chosen to be random states, which are obtained by a Haar random unitary $U_x$ applied to a reference state $\ket{0_x}$. Entanglement properties of the tensor network state $\ket{\Psi}$ can be simplified if we average over the choice of $\ket{V}$, because the ensemble of states $\ket{V}\bra{V}$ has a high symmetry. 
However, if we choose $\ket{V_x}$ to be random states, each particular $\ket{\Psi}$ obviously cannot preserve any symmetry of the boundary. In this paper we would like to construct a family of tensor network states that preserve a global symmetry on the boundary. To generalize the idea of RTN, we would like to construct states that are ``completely random except for the symmetry requirement".

We consider a global symmetry group $G$, which could be a Lie group or a finite group. The symmetry transformation acts on each site of the boundary:
\begin{align}
    g\ket{\Psi}=\otimes_{x\in B}g_x\ket{\Psi}
\end{align}
In general, it is possible for the symmetry transformation $G_x$ of each site to be in different representations. 

A natural way to construct a symmetric tensor network state $\ket{\Psi}=\braket{V}{\Psi_P}$ is to require $\ket{\Psi_P}$ and $\ket{V}$ to both be symmetric.\cite{schuch2010peps} We consider a $\ket{\Psi_P}$ in the form of EPR pairs $\ket{\Psi_P}=\otimes_{xy}\ket{xy}$. The symmetry requires that each $\ket{xy}$ is invariant. We denote the Hilbert space of the two ends of this edge as $\mathbb{H}_{xy}$ and $\mathbb{H}_{yx}$. They must have the same dimension and carry conjugate representations, in order for the EPR pair to be invariant. In general, we can decompose $\mathbb{H}_{xy}$ into irreducible representations:
\begin{align}
    \mathbb{H}_{xy}=\oplus_j n_j\mathbb{H}_j
\end{align}
Here $j$ labels the irreps and $n_j$ is the multiplicity of each irrep. \footnote{In the case of Lie group, if the Hilbert space dimension is finite, only a finite subset of irreps have a nonzero multiplicity $n_j$.} Then $\mathbb{H}_{yx}$ must be decomposed as
\begin{align}
    \mathbb{H}_{yx}=\oplus_j n_j\mathbb{H}_{\overline{j}}
\end{align}
We can define a complete basis of $\mathbb{H}_{xy}$ by
\begin{align}
    \ket{j\mu a},~\mu=1,2,...,d_j,~a=1,2,...,n_j
\end{align}
Here $d_j={\rm dim}\kc{\mathbb{H}_j}$ is the dimension of the irrep $j$. For each given $j$, $\mu$ is the index that transforms nontrivially in symmetry transformation, while $a$ transforms trivially. Without losing generality, we can define the EPR pair state as
\begin{align}
    \ket{xy}=\Omega^{-1/2}\sum_{j,\mu,a}\ket{j\mu a}_{xy}\ket{\overline{j},\mu,a}_{yx}\label{eq:SRTN_edge}
\end{align}
with $\Omega=\sum_jd_jn_j$ a normalization constant. The symmetry allows a nontrivial off-diagonal wavefunction in $a$ index, such as $\sum_{j,\mu,a,b}\phi_{ab}\ket{j\mu a}_{xy}\ket{\overline{j}\mu b}_{yx}$, but that can always be absorbed to the definition of vertex tensors, so we choose the simplest expression above for the link state $\ket{xy}$.

Then the requirement of $\ket{V}$ being symmetric implies that each vertex state $\ket{V_x}$ is symmetric. $\ket{V_x}$ is defined in the Hilbert space $\mathbb{H}_x=\otimes_{y{\rm~nn~}x}\mathbb{H}_{xy}$, which is a product of all qubits $xy$ defined at the $x$-end of each edge $xy$ connected with $x$. The symmetry requirement of $\ket{V_x}$ simply means that the representation carried by all edges fuse to a trivial representation. Without losing generality, let's assume we have a trivalent graph, {\it i.e.} each vertex is connected with three edges. In this case the vertex tensor has the form
\begin{align}
    \ket{V_x}=\sum_{jkl,\mu\nu\sigma,abc}R_{abc}^{jkl}C^{jkl}_{\mu\nu\sigma}\ket{j\mu a}_{xy_1}\ket{k\nu b}_{xy_2}\ket{l\sigma c}_{xy_3}\label{eq:SRTN_vertex}
\end{align}
The $jkl$ representations are required to fuse to a trivial representation, which means the $\mu\nu\sigma$-dependent term is completely determined by the Clebsch-Gordan coefficients $C_{\mu\nu\sigma}^{jkl}$. The muplicity-dependent term $R_{abc}^{jkl}$ can be arbitrary, which in general also depends on $j,k,l$. 

Eq. (\ref{eq:SRTN_edge}) and (\ref{eq:SRTN_vertex}) defines a general symmetric tensor network. Now we would like to generalize the story of random tensor networks, and consider an ensemble of such tensor networks. The simplest choice of ensemble is to assume $R_{abc}^{jkl}$ as a random tensor of $abc$ indices for each given $jkl$, which are independent for different $jkl$:
\begin{align}
R_{abc}^{jkl}&=\lambda^{jkl}T_{abc}^{jkl}
\end{align}
with $T_{abc}^{jkl}$ a random vector of index $abc$ (with dimension $n_jn_kn_l$), normalized as $\sum_{abc}\abs{T_{abc}^{jkl}}^2=1$. $\overline{T_{abc}^{jkl}T_{a'b'c'}^{j'k'l'}}=0$ unless $j=j',k=k',l=l'$. The coefficients $\lambda^{jkl}$ are tunable parameters in this ensemble. 

\subsection{Emergent gauge field}

\begin{figure}
    \centering
    \includegraphics[width=5.5in]{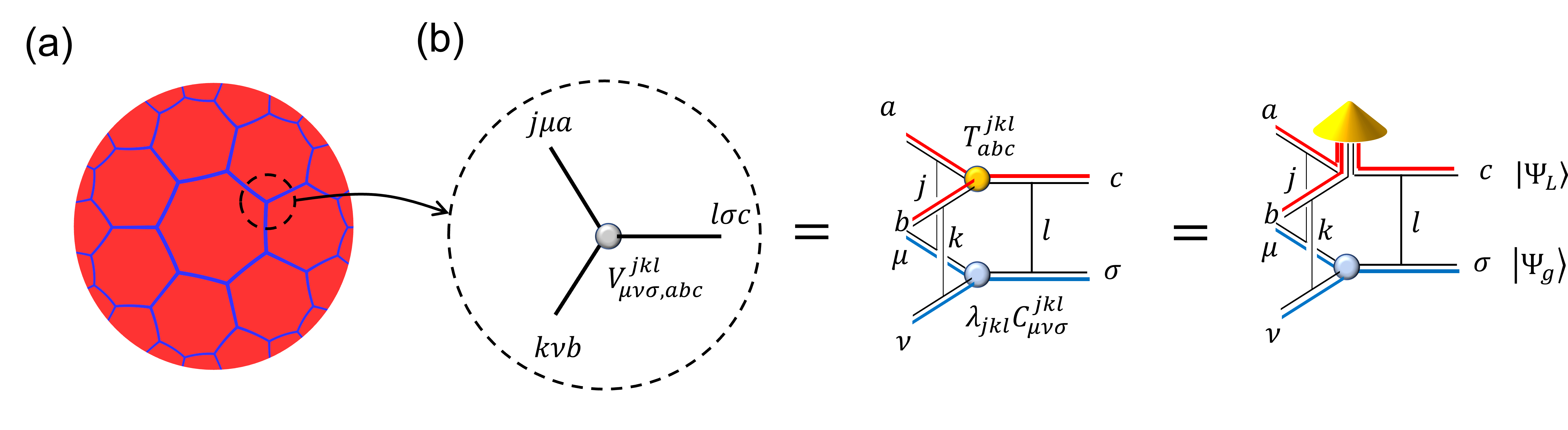}
    \caption{(a) Illustration of a tensor network defined on a trivalent graph. (b) Each tensor is a three-qudit state with the Hilbert space of each qudit decomposed into irreducible representations of the symmetry group. (b) Illustration of the decomposition of the symmetric random tensor network. The vertex tensor is defined to be a product of Clebsch-Gordan coefficients $C_{\mu\nu\sigma}^{jkl}$, a normalized random tensor $T_{abc}^{jkl}$ that is independent for different representations and different vertices, and a weight $\lambda^{jkl}$. The tensor network state can be expressed as the vertex projection by random tensors $T_{abc}^{jkl}$ on a state $\ket{\tilde{\Psi}_P}=\ket{\Psi_L}\otimes\ket{\Psi_g}$, with $\ket{\Psi_g}$ defining the bulk gauge field wavefunction in the basis of representations. The yellow ball and the yellow cone both refers to the projection to the random state $\ket{T_x}$ defined by tensor $T_{abc}^{jkl}$. }
    \label{fig:setup}
\end{figure}
To understand entanglement behavior of the state defined above, we can rewrite the state as
\begin{align}
    \ket{\Psi}&=\braket{T}{\tilde{\Psi}_P}\\
    \ket{\tilde{\Psi}_P}&=\hat{E}_L\ket{\Psi_g}\label{eq:PsitildeP}
\end{align}
Here $\ket{T}=\otimes \ket{T_x}$ is a product of random vertex states $\ket{T_x}=\sum_{abc,jkl}T_{abc}^{jkl}\ket{ja}_{xy_1}\ket{kb}_{xy_2}\ket{lc}_{xy_3}$, with $y_1,y_2,y_3$ the three neighbors of $x$. Note that the Hilbert space here is not the same as original $\mathbb{H}_{xy}$, since we have got rid of the indices $\mu$ that transforms under symmetry. $\ket{T_x}$ is defined in a smaller Hilbert space $\tilde{\mathbb{H}}_{x}=\otimes_{y{~{\rm nn}~}x}\tilde{\mathbb{H}}_{xy}$, with ${\rm dim}\kc{\tilde{\mathbb{H}}_{xy}}=\sum_jn_j$. 

$\ket{\Psi_g}$ is a state obtained by contracting $\mu\nu\sigma$ indices of the tensors $\lambda^{jkl}C_{\mu\nu\sigma}^{jkl}$ (see Fig. \ref{fig:setup} (b)). $\ket{\Psi_g}$ lives in a Hilbert space that is much smaller than the original $\ket{\Psi_P}$, spanned by representation index $j$ on each link. 
On a dangling edge connected with boundary, the states are defined as $\sket{j\mu}$ with $\mu$ labelling states in the irreducible representation $j$. The contraction of tensors $\lambda^{jkl}C_{\mu\nu\sigma}^{jkl}$ gives a wavefunction of $\ket{\Psi_g}$ of the form
\begin{align}
    \ket{\Psi_g}=\sum_{\ke{j_{xy}},\ke{\mu_x,~x\in B}}\Psi_g\kc{\ke{j_{xy}},\ke{\mu_x,~x\in B}}\otimes_{\overline{xy}\in I}\sket{j_{xy}}\otimes \otimes_{x\in B,y\in I}\sket{j_{xy}\mu_x}\label{eq:Psi_g}
\end{align}
which has $j_{xy}$ as the bulk degrees of freedom and $j_{xy}\mu_x$ as the boundary degrees of freedom. 
$\ket{\Psi_g}$ is what we will identify as the wavefunction of a lattice gauge theory. (Tensor network states have been used to study lattice gauge theory. See {\it e.g.} \cite{tagliacozzo2014tensor,haegeman2015gauging}) Before discussing more properties of $\ket{\Psi_g}$ in Sec. \ref{sec:gauge}, we would like to discuss entanglement properties of this symmetric RTN state.

$\hat{E}_L$ is an isometry which embeds the state $\sket{j}$ to the EPR pair state in the corresponding multiplicity space:
\begin{align}
    \hat{E}_L&=\prod_{\overline{xy}}\hat{E}_{Lxy}\nn\\
    \hat{E}_{Lxy}&=\sum_j\frac1{\sqrt{n_j}}\ket{ja}_x\ket{\overline{j}a}_y{~}_{xy}\sbra{j}
\end{align}


Since $\ket{T}=\otimes_x\ket{T_x}$ is now a ``standard" random tensor just like RTN without symmetry, we can apply the method of random averaging and compute entanglement properties of $\ket{\Psi}$. We would like to consider the limit that the multiplicity space dimension $n_j$ is large, which means dominant entanglement entropy is contributed by $\ket{\Psi_L}$ (red lines in Fig. \ref{fig:setup}). 

\subsection{Second Renyi entropy}

As a simplest example of entanglement features, we consider the computation of purity and second Renyi entropy. For a given quantum state $\rho$, the purity of a subsystem $A$ can be expressed as $p_A={\rm tr}\kc{\rho_A^2}={\rm tr}\kc{\rho\otimes \rho X_A}$. Here $X_A$ is the swap operator that exchanges two replica of $A$. Picking any complete basis of $A$ and $\overline{A}$ as $\ket{n_A}\ket{s_{\overline{A}}}$, $X_A$ is defined as
\begin{align}
    X_A=\sum_{n,m}\ket{n_A}_1\ket{m_A}_2{}\leftidx{_1}{\bra{m_A}}\leftidx{_2}{\bra{n_A}}
\end{align}
The second Renyi entropy is defined as $S_A^{(2)}=-\log p_A$. For the ensemble of symmetric random tensor network states defined above, we can define the averaged purity of a boundary region $A$ as
\begin{align}
    p_A=\frac{{\rm tr}_B\kd{\overline{\rho\otimes\rho} X_A}}{{\rm tr}_B\kd{\overline{\rho\otimes \rho}}}\label{eq:purity}
\end{align}
Here $\rho=\ket{\Psi}\bra{\Psi}=\braket{T}{\tilde{\Psi}_P}\braket{\tilde{\Psi}_P}{T}$, and the denominator is a normalization since the RTN is not normalized. Note that the average over states is carried before normalizing the state. This is in general different from the average after normalization, but we will focus on the large bond dimension limit, in which case the difference is suppressed\cite{hayden2016holographic}. 

The numerator can be expressed as
\begin{align}
    Z_A\equiv {\rm tr}_B\kd{\overline{\rho\otimes\rho} X_A}=\bra{\tilde{\Psi}_P}^{\otimes 2}X_A\otimes \otimes_{x\in I}\overline{\kc{\ket{T_x}\bra{T_x}}^{\otimes 2}}\ket{\tilde{\Psi}_P}^{\otimes 2}
\end{align}
The average of random state $\ket{T_x}$ is known as
\begin{align}
    \overline{\kc{\ket{T_x}\bra{T_x}}^{\otimes 2}}=\frac1{D_x(D_x+1)}\kc{X_x+\mathbb{I}_x}
\end{align}
which allows $Z_A$ to be expressed as a sum over subsets of the bulk:
\begin{align}
    Z_A&=C^{-1}\sum_{\Omega\subseteq I}e^{-S_{\Omega A}^{(2)}\kc{\rho_P}}\label{eq:generalS2}
\end{align}
Here the second Renyi entropy on the right-hand side is defined for $\rho_P=\ket{\tilde{\Psi}_P}\bra{\tilde{\Psi}_P}$ for the subsystem $\Omega\cup A$. The denominator in Eq. (\ref{eq:purity}) can be expressed in the same way by replacing $A$ by the emptyset $\emptyset$. Thus we obtain
\begin{align}
    p_A=\frac{Z_A}{Z_\emptyset}\label{eq:averaged purity}
\end{align}
The derivation of Eq. (\ref{eq:generalS2}) and (\ref{eq:averaged purity}) applies to general $\rho_P$. In a simplest RTN, when $\rho_P$ consists of maximally entangled EPR pairs with dimension $D$, $S_{\Omega A}^{(2)}\kc{\rho_P}=\abs{\partial(\Omega A)}\log D$ is proportional to the area of the domain wall separating $\Omega$ and $A$ with the complement. In the large bond dimension limit, the sum in Eq. (\ref{eq:generalS2}) is dominated by the largest term, which reduces to the RT formula $S_A^{(2)}\simeq {\rm min}_{\Omega}\log D\abs{\pa\kc{\Omega A}}$. 

The state $\ket{\tilde{\Psi}_P}$ defined in Eq. (\ref{eq:PsitildeP}) is not a product of EPR pairs. Now we study the purity of this state. It should be noted that the link variable $j$ is copied to the two neighboring sites. In other words, the state $\sket{j}_{xy}$ is viewed as $\sket{j}_{xy}\otimes\sket{j}_{yx}$ where $\sket{j}_{xy}$ is in the Hilbert space at site $x$, while $\sket{j}_{yx}$ is in that of site $y$. In this way, we can still define partition of the system by a subset of vertices $\Omega$. For a given $\Omega$, tracing over the complement $\overline{\Omega}$ will require the density operator to be diagonal in $j$ for all boundary links. Denote the boundary $\partial\kc{\Omega A}$ as $\gamma$, and denote a configuration of $j_{xy}$ intersecting with the boundary $\gamma$ by $J=\ke{j_{xy},x\in \Omega,y\in \overline{\Omega}}$, we can define a projection operator
\begin{align}
    P_J=\otimes_{x\in\Omega,y\in\overline{\Omega}}\sket{j_{xy}}\sbra{j_{xy}}
\end{align}
One can prove that the reduced density operator has the following decomposition:
\begin{align}
\rho_{\Omega A}^P&\equiv {\rm tr}_{\overline{\Omega}\overline{A}}\kc{\ket{\tilde{\Psi}_P}\bra{\tilde{\Psi}_P}}=\sum_JP_J{\rm tr}_{\overline{\Omega}\overline{A}}\kc{\ket{\Psi_g}\bra{\Psi_g}}P_J\otimes \otimes_{j_{xy}\in J}\frac{\mathbb{I}_{xyj}}{n_{j_{xy}}}\label{eq:rdm_decomposition}
\end{align}
Here $\frac{\mathbb{I}_{xyj}}{n_{j_{xy}}}$ is the maximally mixed density matrix in the multiplicity space of link $\overline{xy}$. Note that the isometry $\hat{E}_{Lxy}$ only has a nontrivial effect to the entanglement entropy on links that intersects with the boundary $\gamma$. On the right-hand side of Eq. (\ref{eq:rdm_decomposition}), the partial trace is carried in the smaller Hilbert space of the gauge field. Denote
\begin{align}
    p_J&={\rm tr}\kd{P_J{\rm tr}_{\overline{\Omega}\overline{A}}\kc{\ket{\Psi_g}\bra{\Psi_g}}}\\
    \sigma_{gJ}&=p_J^{-1}P_J{\rm tr}_{\overline{\Omega}\overline{A}}\kc{\ket{\Psi_g}\bra{\Psi_g}}P_J
\end{align}
Eq. (\ref{eq:rdm_decomposition}) implies that
\begin{align}
   {\rm tr}\kc{{\rho_{\Omega A}^{P}}^2}=\sum_J\frac{p_J^2}{n_J}{\rm tr}\kc{\sigma_{gJ}^2},~n_J=\prod_{xy\cap \gamma}n_{j_{xy}}
\end{align}
Compared with the simple random tensor network, $n_J^{-1}$ corresponds to $D^{-\abs{\gamma}}=e^{-\log D\abs{\gamma}}$ which contributes the area law term. ${p_J^2}{\rm tr}\kc{\sigma_{gJ}^2}$ is the contribution from the gauge field. 

We are interested in the case when the area law contribution $-\log n_J$ is dominant compared with the gauge field contribution. As a simplest case, we can take $n_j=D$ for all $j$, which means each representation has the same multiplicity. In that case, the expression simplifies to
\begin{align}
    {\rm tr}\kc{{\rho_{\Omega A}^{P}}^2}&=e^{-\abs{\gamma}\log D}\sum_Jp_J^2{\rm tr}\kc{\sigma_{gJ}^2}\equiv e^{-\abs{\gamma}\log D-S^{g(2)}_{\Omega A}}
\end{align}
The second term $S^{g(2)}_{\Omega A}=-\log {\rm tr}_{\overline{\Omega}\overline{A}}\kc{\ket{\Psi_g}\bra{\Psi_g}}$ is the second Renyi entropy of the gauge field state. If there is one choice of $\Omega$ dominating the sum in Eq. (\ref{eq:generalS2}), we obtain the following quantum extremal surface formula:
\begin{align}
    S_A^{(2)}\simeq \min_{\Omega}\kc{\log D\abs{\pa(\Omega A)}+S_{\Omega A}^{g(2)}}\label{eq:RT with gauge}
\end{align}
In particular, if we take the limit of large $D$ with $S^{g(2)}_{\Omega A}$ finite, and if the minimal surface $\gamma=\partial\kc{\Omega A}$ is unique, the sum in the partition function is indeed dominated by one term, and the approximation holds. 

Eq. (\ref{eq:RT with gauge}) can be viewed as a confirmation that there is indeed a gauge field described by state $\ket{\Psi_g}$ living in the bulk. If we consider the more general case with $n_j$ depending on $j$, but assume that for all $j$ $n_j$ is large, then the physical picture qualitatively remains the same, except that the geometry ({\it i.e.} the area law contribution to entropy) is correlated with the gauge field configuration. 

The discussion here can be generalized to higher Renyi entropies in the same way as the original RTN. In the case $n_j=D$ for all $j$, in the large $D$ limit we have
\begin{align}
    S_A^{(n)}\simeq \min_{\Omega}\kc{\log D\abs{\pa(\Omega A)}+S_{\Omega A}^{g(n)}}\label{eq:Sn with gauge}
\end{align}
In the same limit, one can prove that fluctuations of the Renyi entropies are subleading in $\frac1D$, so that the Renyi entropy of each realization of the random tensor network is close to the random average value. Because these results directly follow from the results in Ref. \cite{hayden2016holographic}, we won't go to further details in this paper.

\section{More analysis on the bulk gauge field state}\label{sec:gauge}

\subsection{An overview of the lattice gauge theory}
\label{subsec:lattice gauge}

In this subsection, we will elaborate more on the interpretation of the state $\ket{\Psi_g}$ as a lattice gauge theory state. We will start by an overview of the lattice gauge theory. 

We start from a continuous Yang-Mills theory with the action
\begin{align}
    S&=-\frac1{4g}\int d^dx{\rm tr}\kc{F_{\mu\nu}F^{\mu\nu}}\\
    F_{\mu\nu}&=\pa_\mu A_\nu-\pa_\nu A_\mu+i\kd{A_\mu,A_\nu}\nn\\
    A_\mu&=A_\mu^aT_a
\end{align}
Here $A_\mu$ takes its value in the Lie algebra, with $T_a$ the generators. For simplicity we will for now discuss the action in flat space, although for the purpose of our work we are actually interested in this theory in curved space. 
We can pick a gauge $A_0=0$, in which case the electric field is $E_{i}=F_{0i}=\pa_0A_i$. In this gauge we can define the canonical quantization
\begin{align}
S&=\frac1{2g}\int d^{d-1}xdt{\rm tr}\kd{E_i^2-\frac12F_{ij}F^{ij}}\\
H&=\int d^{d-1}x{\rm tr}\kd{\frac{g}2\Pi_i^2-\frac1{4g}F_{ij}F^{ij}}\\
    \Pi_i&=\frac{\delta\mathcal{L}}{\delta \pa_0A_i}=\frac1g\pa_0A_i
\end{align}
Now we consider the lattice regularization in spatial direction, while keeping the time direction continuous\cite{kogut1975hamiltonian}. The discretization is done by choosing a graph (for example a cubic lattice) in space, and define a group element $u_{xy}\in G$ for each oriented link $xy$. We also define $u_{yx}\equiv u_{xy}^{-1}$. Approximately, when we compare the lattice theory with the continuous theory, $u_{xy}=P\exp\kd{i\int_x^y A_i(z) dz^i}$ corresponds to the Wilson line operator of the continuous theory. Gauge transformation is defined as
\begin{align}
    u_{xy}\>g_x^{-1}u_{xy}g_y
\end{align}
for arbitrary $g_x,g_y\in G$. The gauge invariant magnetic flux is defined for each closed loop $L=\ke{x_1x_2...x_nx_1}$:
\begin{align}
    \Phi(L)={\rm tr}\kd{u_{x_1x_2}u_{x_2x_3}...u_{x_nx_1}}\label{eq:Wilson loop}
\end{align}
$\Phi(L)$ is called the Wilson loop operator, which is the non-Abelian version of the (exponential of) net magentic flux threaded into a loop $L$. Picking a set of fundamental loops of the graph (such as squares of a cubic lattice), and denote them as $L_I$, the lattice version of the magnetic field term in the Hamiltonian is
\begin{align}
    H_B=-\frac1{g}\sum_I{\rm Re}{\rm tr}\kd{u_{x_1x_2}...u_{x_nx_1}}\equiv -\frac1{2 g}\sum_I\kc{\Phi\kc{L_I}+\Phi^\dagger\kc{L_I}}
\end{align}
Since $u_{xy}$ are rotors taking value in the group manifold, the conjugate variable should be the ``angular momentum" operator of the rotor. More precisely, the $u_{xy}$ transforms in group $G\times G$ as $u_{xy}\>g_x^{-1}u_{xy}g_y$, so there are two kinds of ``angular momentum" operators that rotate $u_{xy}$, which acts by left multiplication and right multiplication, respectively. We will call them as left translation and right translation. Denote the generators of the left and right translation as $\Pi_L$ and $\Pi_R$, respectively, with $\Pi_L=\Pi_L^aT_a,~\Pi_R=\Pi_R^aT_a$. When we consider the quantum mechanics of the rotor $u_{xy}$, we have
\begin{align}
    \hat{\Pi}_L^a\psi\kc{u}&=-\frac{i}{\epsilon}\kc{\psi\kc{u+\epsilon T_a u}-\psi\kc{u}}=-i\frac{\pa\psi(u)}{\pa u_{\alpha\beta}}T_{a\alpha\gamma}u_{\gamma\beta}\\
\hat{\Pi}_{L}\psi(u)&=-iu\nabla_u\psi(u)\\
\text{similarly~}\hat{\Pi}_R\psi(u)&=-i\nabla_u \psi(u) u
\end{align}
Here $\nabla_u\psi$ is a matrix defined by $\kd{\nabla_u\psi}_{\alpha\beta}=\frac{\pa\psi}{\pa u_{\beta\alpha}}$. The analog of the electric field term $H_E=\int d^{d-1}x\frac g2{\rm tr}\kc{\Pi_i^2}$ is 
\begin{align}
    H_E=\frac g2\sum_{\overline{xy}}{\rm tr}\kc{\hat{\Pi}_{Lxy}^2}
\end{align}
which is the quadratic Casimir of the left action. It should be noticed that 
\begin{align}
    {\rm tr}\kc{\hat{\Pi}_{Lxy}^2}={\rm tr}\kc{\hat{\Pi}_{Rxy}^2}={\rm tr}\kc{u\nabla_u u\nabla_u}
\end{align}
If we decompose the Hilbert space of each link into irreducible representations of $G\times G$, the equation above tells us that the decomposition look like
\begin{align}
    \mathbb{H}_{xy}=\oplus_j\mathbb{H}_{j}\otimes\mathbb{H}_{\overline{j}}
\end{align}
States in the Hilbert space $\mathbb{H}_{xy}$ can be labeled by $\ket{j_{xy}\mu_{xy}\nu_{yx}}$, with $\mu=1,2,...,d_j$ transforming in representation $j$, and $\nu_{yx}$ transforming in $\overline{j}$. The electric field term ${\rm tr}\kc{\hat{\Pi}_{Lxy}^2}$ is an operator that is diagonal in $j$. \footnote{To illustrate this decomposition, if we consider the case of $G=SU(2)$, we can expand $\psi(u)$ into a polynomial of $u$:
\begin{align}
    \psi(u)=\phi^{\alpha\beta}_1u_{\alpha\beta}+\phi^{\alpha\beta,\gamma\delta}_2u_{\alpha\gamma}u_{\beta\delta}+...
\end{align}
The coefficient $\phi_n^{\alpha_1\alpha_2..\alpha_n,\beta_1\beta_2...\beta_n}$ tranforms in the representation $\ell=\frac n2$. The fact that the left action and the right action share the same representation $\ell$ can be seen from permutation symmetry. $u_{\alpha_1\beta_1}u_{\alpha_2\beta_2}...u_{\alpha_n\beta_n}$ is invariant under simultaneous permutation of the $\alpha$ indices and the $\beta$ indices. This suggests that $\phi_n^{\alpha_1\alpha_2...\alpha_n,\beta_1\beta_2...\beta_n}$ must transform in representation $\ell,\ell$. }

The total Hamiltonian of the lattice gauge theory is 
\begin{align}
    H=H_E+H_B=\frac g2\sum_{\overline{xy}}{\rm tr}\kc{\hat{\Pi}_{Lxy}^2}-\frac1{2 g}\sum_I\kc{\Phi\kc{L_I}+\Phi^\dagger\kc{L_I}}\label{eq:LGT Hamiltonian}
\end{align}

The gauge invariance condition means that $u_{xy}$ for different links are not independent rotor variables. The physical Hilbert space is obtained by taking a quotient of $\otimes_{\overline{xy}}\mathbb{H}_{xy}$ over gauge transformations. The gauge invariance condition is defined as
\begin{align}
    \sum_{y{~\rm nn~}x}\hat{\Pi}_{Lxy}=0
\end{align}
For a trivalent graph, this condition requires that the three representations $j_{xy_1},j_{xy_2},j_{xy_3}$ on the three links starting from $x$ fuse to a trivial representation. This requires the wavefunction to take the form
\begin{align}
    \Phi\kc{\ke{j_{xy},\mu_{xy},\nu_{yx}}}=\prod_xC_{\mu_1\mu_2\mu_3}^{j_1j_2j_3}\Psi_g\kc{\ke{j_{xy}}}
\end{align}
Here $j_1,j_2,j_3$ are the abbreviation of $j_{xy_1},j_{xy_2},j_{xy_3}$ and similarly for $\mu_1,\mu_2,\mu_3$. The $\mu$ dependence of the wavefunction is completely determined by the representation labels $j_{xy}$. In other words, there is an isometry that maps the smaller Hilbert space spanned by states $\sket{j_{xy}}$ to the original Hilbert space $\ket{j_{xy}\mu_{xy}\nu_{yx}}$. If we consider a geometry with boundary, and the boundary condition does not require electric field to vanish at the boundary, then in the discretized model it corresponds to a graph with dangling legs. In that case the wavefunction has the generic form
\begin{align}
    \Phi\kc{\ke{j_{xy},\mu_{xy},\nu_{yx}}}=\prod_{x\in I}C_{\mu_1\mu_2\mu_3}^{j_1j_2j_3}\Psi_g\kc{\ke{j_{xy}},\ke{\mu_x,x\in B}}\label{eq:general gauge WF}
\end{align}
Compare Eq. (\ref{eq:general gauge WF}) with Eq. (\ref{eq:Psi_g}), we see that the RTN wavefunction $\ket{\Psi_g}$ is a special case of a gauge field wavefunction. A general wavefunction in Eq. (\ref{eq:general gauge WF}) cannot be decomposed into a product of the form $\lambda^{jkl}C_{\mu\nu\sigma}^{jkl}$. However, ansatz wavefunction of this form is sufficient to describe confined and deconfined phases of the gauge theory, as we will discuss in the next section.

\subsection{Relation with the Levin-Wen model}

Roughly speaking, gauge theory has two kinds of phases, confined phases and deconfined phases.\cite{wilson1974confinement,kogut1975hamiltonian,fradkin1979phase} The signature of a confined phase is that a pair of probe charges are bounded by a linear force.\cite{tHooft1978phase} 
Roughly speaking, for large $g$ the lattice gauge theory Hamiltonian (\ref{eq:LGT Hamiltonian}) prefers to have a confined ground state since minimizing the electric field term $H_E$ will require each link to have a trivial representation $j=0$. For small $g$ the Hamiltonian prefers a deconfined phase in its ground state since minimizing $H_B$ requires the magnetic flux to vanish in all loops: 
\begin{align}
    u_{x_1x_2}u_{x_2x_3}...u_{x_nx_1}=\mathbb{I},~\forall L
\end{align}
In the extreme limit $g\>0$, the ground state on a sphere is completely determined by the zero flux condition above together with the gauge invariance condition. The zero flux state $\ket{\Psi_g}$ in the reduced Hilbert space spanned by $\sket{j_{xy}}$ is a special case of the Levin-Wen string-net condensation state\cite{levin2005string}. More quantitatively, if we normalize the Clebsch-Gordan coefficients $C^{jkl}_{\mu\nu\tau}$ as
\begin{align}
    \sum_{\mu\nu\tau}\abs{C^{jkl}_{\mu\nu\tau}}^2=\sqrt{d_jd_kd_l}\label{eq:CG normalization}
\end{align}
and take $\lambda_{jkl}=1$, the wavefunction $\ket{\Psi_g}$ obtained by contracting $\mu\nu\tau$ indices is a Levin-Wen wavefunction. This can be verified graphically by reproducing the Levin-Wen conditions. If each vertex is defined as a CG tensor, the following ``$F$-move" equation is automatically satisfied
\begin{align}
   {\includegraphics[scale = 0.3,valign=c]{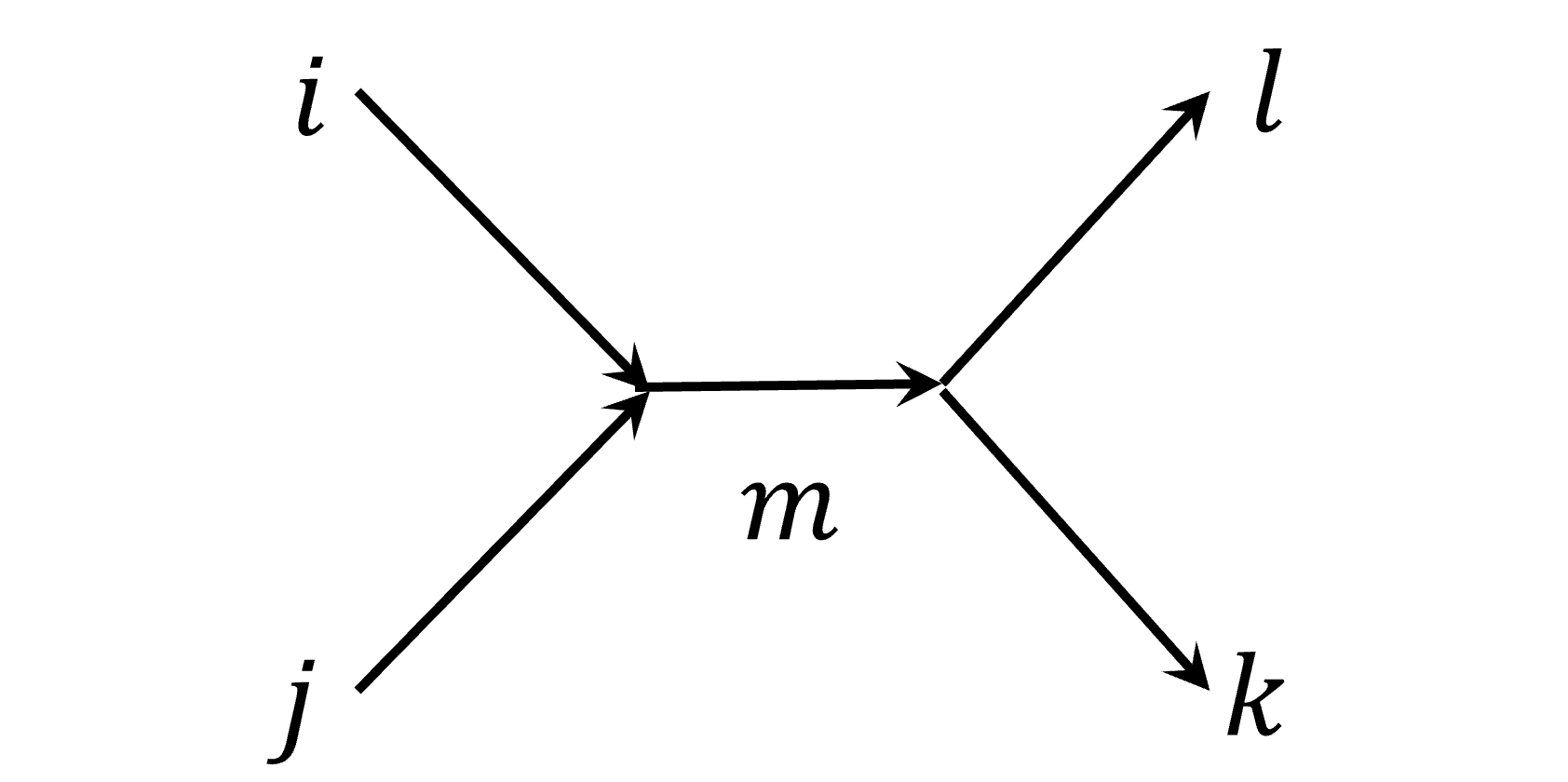}}&=\sum_nF_{\overline{k}\overline{l}n}^{ijm}{\includegraphics[scale = 0.3,valign=c]{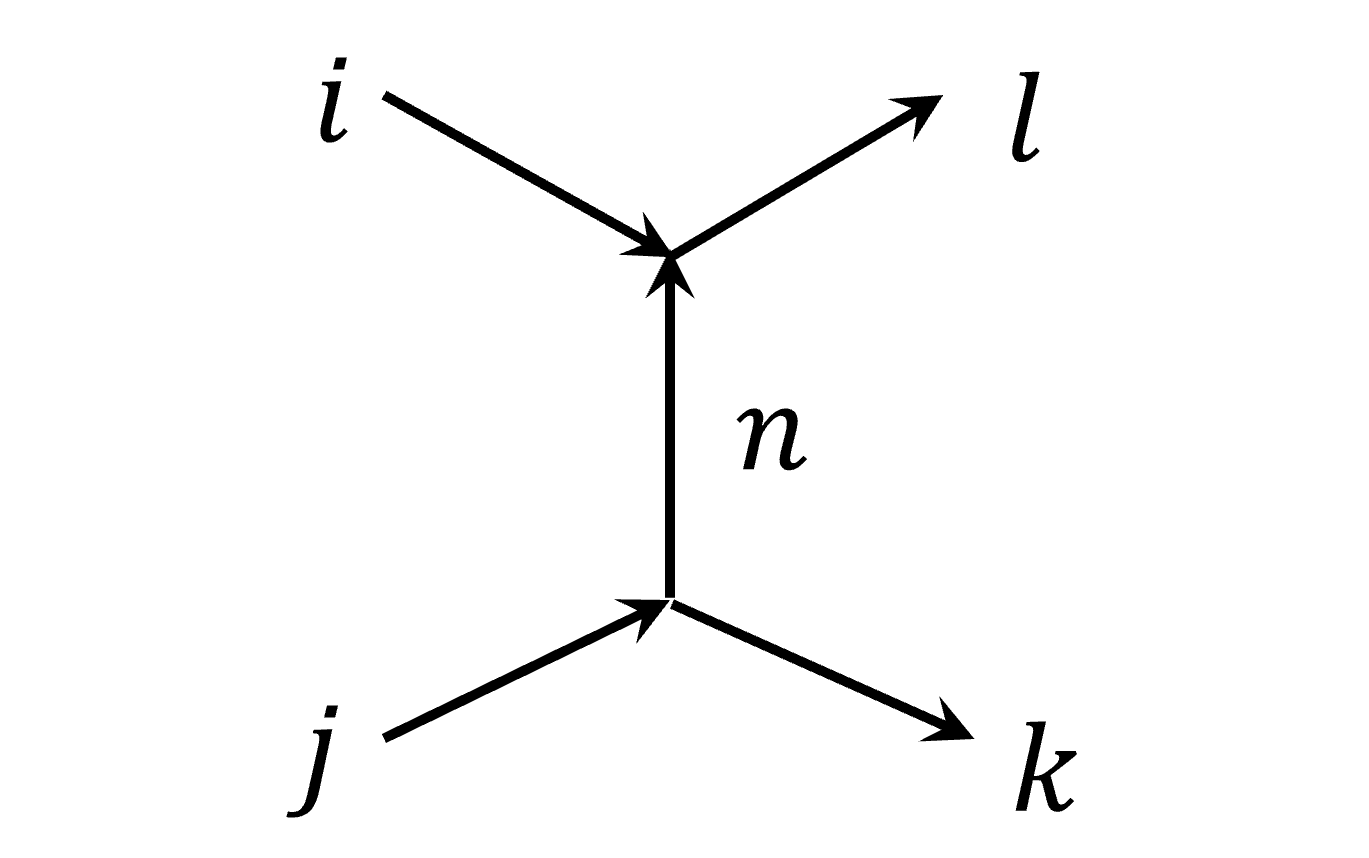}}
\end{align}
by choosing $F_{\overline{k}\overline{l}n}^{ijm}$ to be the $6j$ symbol. (In the drawing we didn't write the $\mu\nu$ indices explicitly.) The condition that determines the normalization of the CG coefficient is
\begin{align}
    \includegraphics[scale=0.3,valign=c]{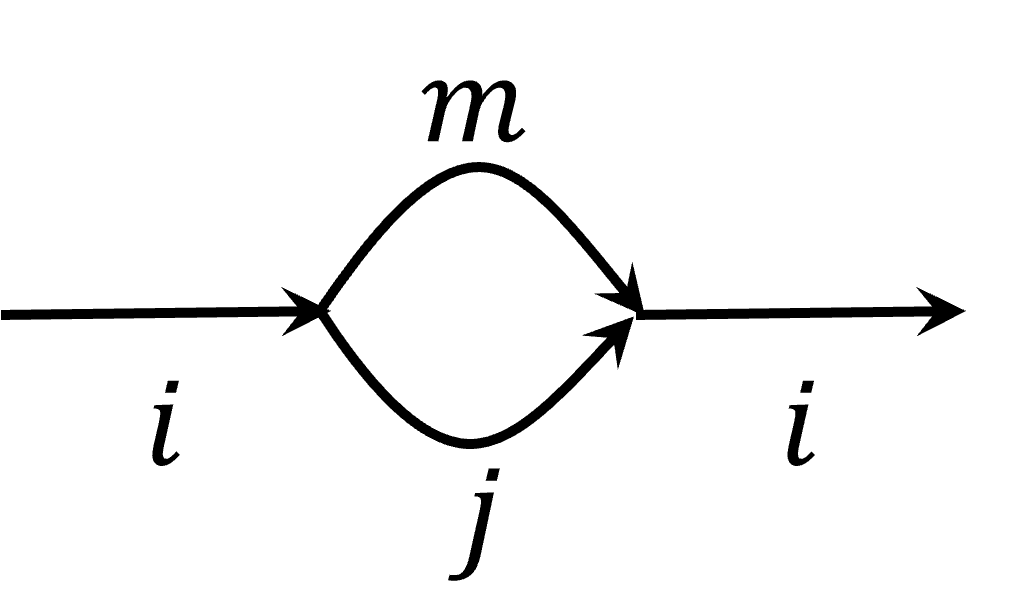}=F_{\overline{j}\overline{i}0}^{ijm}\includegraphics[scale=0.3,valign=c]{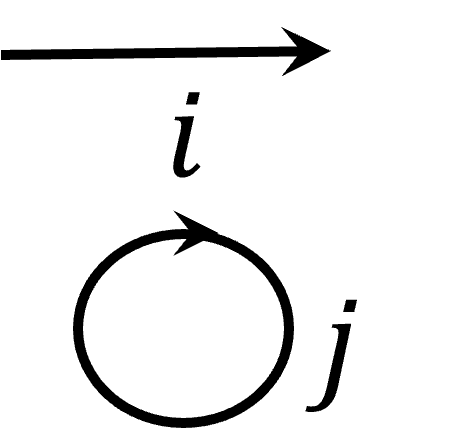}=\sqrt{\frac{d_jd_m}{d_i}}\includegraphics[scale=0.3,valign=c]{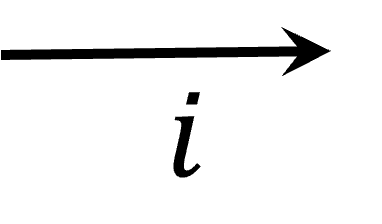}
\end{align}
It should be remembered that the horizontal line represents a $\delta$ function in $\mu=1,2,...,d_i$. Taking a sum over $\mu$ leads to Eq. (\ref{eq:CG normalization}).\footnote{We thank Michael Levin for pointing out this fact to us.}

This result explicitly tell us that the wavefunction $\ket{\Psi_g}$ in Eq. (\ref{eq:Psi_g}) can describe a gauge field theory ground state in the (extremely) deconfined phase. On the other hand, one can obviously tune the coefficient $\lambda_{jkl}$ to make a confined phase, by requiring that $\lambda_{jkl}\ll \lambda_{000}$ as long as some of $j,k,l$ are not the trivial representation $0$. For example, we can define $\lambda_{jkl}=p_jp_kp_l$ with $p_k<p_0$ for all $k\neq 0$. This corresponds to a ``line tension" that depends on $k$. Thus we have shown that this family of symmetric random tensor networks can describe emergent bulk gauge field with deconfined or confined phases. 

It should be noted that for a Lie group gauge theory, the Hilbert space for $\lambda_{jkl}=1$ state is infinite on each link. More general Levin-Wen models are also tensor network states\cite{buerschaper2009explicit}. It is possible to define some ensemble of random tensor networks with a more general Levin-Wen model in the bulk, but we won't discuss that further in this paper.

\subsection{Topological entropy}

In the previous subsection, we have shown that our ensemble of symmetric RTN can realize the confined phase and deconfined phase of the gauge field. According to the quantum extremal surface formula (\ref{eq:RT with gauge}) and (\ref{eq:Sn with gauge}), the gauge field contribution to the Renyi entanglement entropy of a boundary region $A$ is $S_{\Omega A}^{g(n)}$, which is the Renyi entropy of region $\Omega A$ for the gauge field state $\ket{\Psi_g}$. $\Omega$ is the region between the minimal surface and region $A$, which is called (a spacial slice of) the entanglement wedge of $A$. A natural question is the boundary consequence of the bulk phase transition between confined and deconfined phases of the gauge field. 

To address this question, we consider the topological entropy\cite{levin2006detecting,kitaev2006topological}, which is a feature of topologically ordered states, such as Levin-Wen models. The entanglement entropy of a subsystem $\Sigma$ in a Levin-Wen state has the form
\begin{align}
    S_\Sigma=\alpha\abs{\pa\Sigma}+S_{\rm topo}
\end{align}
The coefficient of area law $\alpha$ is not universal, but the constant correction $S_{\rm topo}$ is universal and negative. For the Levin-Wen state it is determined by the total quantum dimension $D_T$:
\begin{align}
    S_{\rm topo}=-\log D_T,~D_T=\sum_id_i^2\label{eq:topo entropy}
\end{align}
For example for $Z_n$ gauge theory, we have $n$ irreducible representations each with dimension $1$, so that $D_T=n$. \footnote{For generic topologically ordered states, we need to define the quantum dimension for each particle type $d_a$, which is not necessarily an integer. The quantum dimension is $D_T=\sqrt{\sum_ad_a^2}$. Note that the formula is different from Eq. (\ref{eq:topo entropy}) because the Levin-Wen model is a quantum double of the tensor category used to define the types of string-nets.} 

\begin{figure}
    \centering
    \includegraphics[width=3.5in]{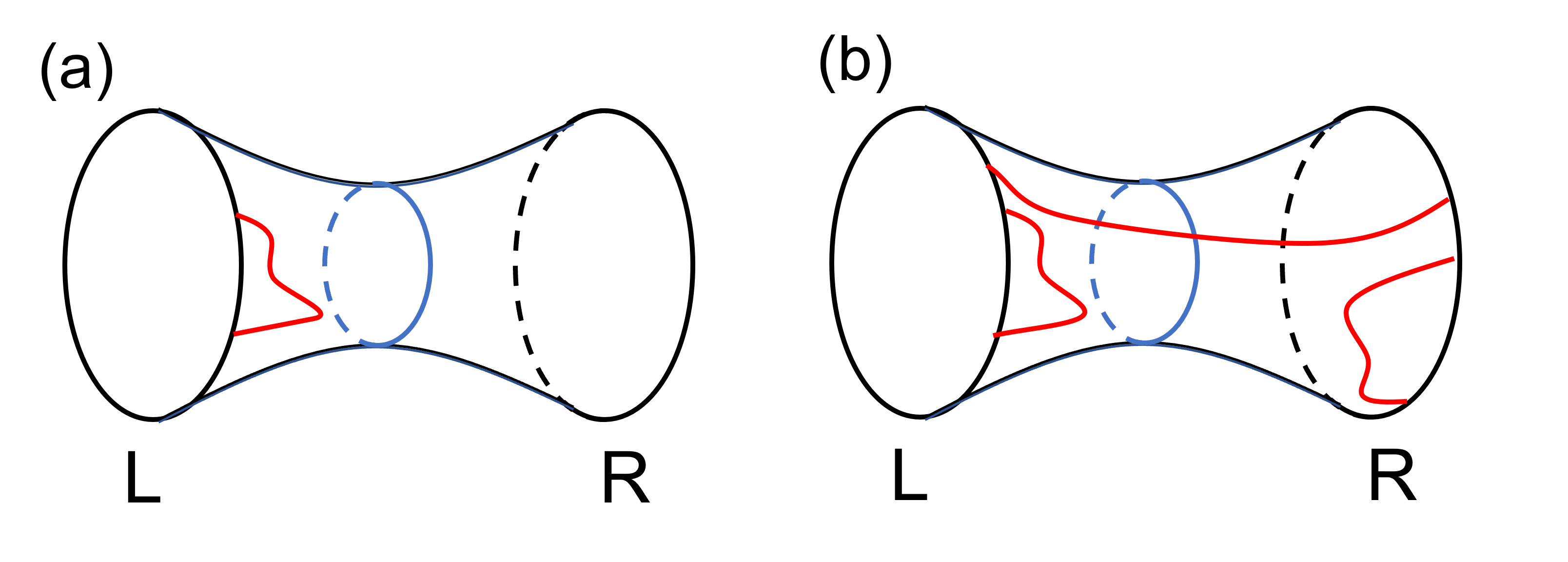}
    \caption{The wormhole geometry corresponding to two entangled boundaries in a TFD state. (a) and (b) illustrate the two topological sectors in the case of $Z_2$ gauge theory. The blue circle is the black hole horizon, which is also the RT surface of a single boundary.}
    \label{fig:TFD}
\end{figure}

Now we would like to understand the boundary interpretation of the topological entropy. For this purpose, we consider the eternal black hole geometry in $2+1$-dimensions, in which case the spatial geometry is a wormhole connecting two boundaries\cite{maldacena2003eternal}. The corresponding state on the boundary is known to be a thermofield double (TFD) state:
\begin{align}
    \ket{TFD}=Z^{-1/2}\sum_ne^{-\beta E_n/2}\ket{n}_L\ket{n}_R
\end{align}
$E_n$ are energy eigenvalues of the boundary theory, and the TFD state is a purification of the thermal ensemble. Now we consider the symmetric random tensor network living on a graph that is a discretization of this wormhole geometry, as is shown in Fig. \ref{fig:TFD}. This geometry has a non-contractable loop, which allows different topological sectors of the bulk gauge theory. For example, a $Z_2$ gauge theory in the deconfined phase has two topological sectors shown in Fig.\ref{fig:TFD}, which correspond to charge $+1$ and $-1$ on each boundary. when we fix the topological sector, this corresponds to fixing the charge on the boundary, which means we are considering a thermofield double state that has a definite charge---{\it i.e.} a canonical ensemble instead of a macrocanonical ensemble:
\begin{align}
    \ket{TFD_j}=Z^{-1/2}\sum_{q_n=j}e^{-\beta E_n/2}\ket{n}_L\ket{n}_R
\end{align}
$\ket{TFD_j}$ is labeled by an irreducible representation $j$. In large $D$ limit, the entropy of one boundary in this state is determined by the RT formula with quantum correction:
\begin{align}
    S_R=\log D\abs{A_{BH}}+S_{\Sigma_RR}^{g}
\end{align}
Here $\abs{A_{BH}}$ is the area of the black hole horizon, and $\Sigma_R$ is the entanglement wedge of the right boundary, which is the region between $R$ and the black hole horizon. If the gauge field is in a topological state, for $\ket{TFD_0}$ with trivial representation, $S_{\Sigma_RR}^{g}=\alpha\abs{A_{BH}}-\log D_T$, so that
\begin{align}
     S_R=\kc{\log D+\alpha}\abs{A_{BH}}-\log D_T
\end{align}

\begin{figure}
    \centering
    \includegraphics[width=3.5in]{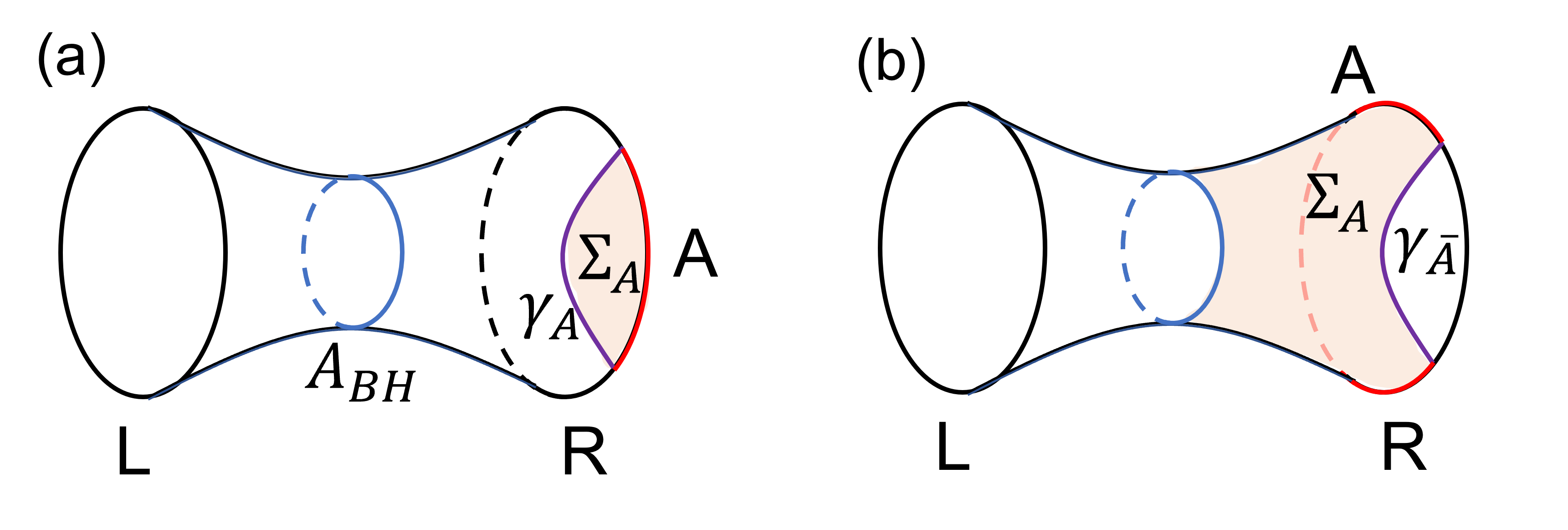}
    \caption{(a) A small subsystem $A$ (red arc) with a singly connected entanglement wedge $\Sigma_A$ (orange shaded region). (b) A large subsystem $A$ (red arc) with a doubly connected entanglement wedge $\Sigma_A$. The RT surface $\gamma_A=\gamma_{\overline{A}}\cup A_{BH}$ with $A_{BH}$ the black hole horizon. }
    \label{fig:TFD2}
\end{figure}

More generally, for $\ket{TFD_j}$, the topological entropy is modified to $\log \frac{D_T}{d_i}$. For the boundary state, the constant correction $-\log D_T$ represents the information about global charge of the boundary. Compared with a macrocanonical ensemble with uncertain total charge, the canonical ensemble has a fixed charge and thus the entropy is reduced by $\log D_T$. We can further consider the entropy of a subsystem, as is illustrated in Fig. \ref{fig:TFD2}. Upon increasing the size of the subsystem, there is a phase transition in the topology of the entanglement wedge $\Sigma_A$. For a small region, $\Sigma_A$ is singly connected. For a large region $\Sigma_A$ encloses the black hole and its boundary $\gamma_A=A_{BH}\cup \gamma_{\overline{A}}$ has two connected components. The entropy of $A$ is given by the minimum of the two configurations:
\begin{align}
    S_A={\rm min}\ke{\abs{\gamma_A}\kc{\log D+\alpha},\kc{\abs{A_{BH}}+\abs{\gamma_{\overline{A}}}}\kc{\log D+\alpha}-\log D_T}
\end{align}
The constant term $-\log D_T$ only occurs in the non-singly connected phase, since the charge across the open curve $\gamma_A$ or $\gamma_{\overline{A}}$ is not fixed. This constant shift for large region suggests that a large region $A$ knows about the net charge of the entire boundary, while a small region $A$ in the singly connected phase has no knowledge about the net charge.

On comparison, if the bulk gauge field is in confined phase, there is no topological entropy, and the entropy of a boundary region satisfies RT formula without constant correction, independent from the topology of the entanglement wedge. The confined phase corresponds to an ``insulator" phase on the boundary, where the system is not only neutral but locally neutral: charge fluctuations are restricted to short-strings near the boundary of the region. In the wormhole geometry (\ref{fig:TFD}), in the confined phase the charge on each boundary will be trivial. More precisely, the probability of having a nontrivial charge is exponentially small: $p_{j\neq 0}\sim e^{-d_{LR}/\xi}$ with $d_{LR}$ the distance between the two boundaries, and $\xi$ a confinement length scale. Therefore the entanglement entropy of the black hole $S_R$ satisfies an area law $S_R=\kc{\log D+\alpha}\abs{A_{BH}}$ independent from the charge boundary condition.

\section{Conclusion and discussions}\label{sec:conclusion}

In conclusion, we have studied a family of random tensor networks that preserve a given global symmetry group $G$ on the boundary. We show that the Renyi entropy of this family of RTN agrees with a bulk interpretation of a $G$-gauge theory living on the background bulk geometry. Parameters in this ensemble can be used to tune this gauge theory into a confined phase or a deconfined phase. We show that topological entropy of the bulk gauge theory in the deconfined phase corresponds to the information about global charge (the representation of symmetry group $G$) of the boundary theory. In an wormhole geometry, a large enough region knows about the net charge of the entire boundary, which leads to a reduction of its entropy. The confined phase corresponds to an insulator phase on the boundary where the charge is almost trivial everywhere, and there is no entropy reduction due to knowing the charge. 

There are many open questions about understanding the global-gauge duality in tensor networks. Besides the deconfined phase and confined phase (in which two charged particles have short-range and linear interaction, respectively), for some gauge field theory it is possible to have Coulomb phase. This is a phase with gapless photons and power law interaction between charges. Based on results in AdS/CFT, this should correspond to a metal phase on the boundary\cite{hartnoll2018holographic}. Another question is in the topological phase, whether there are other entanglement properties that can probe the topological entropy directly. For example Ref.\cite{kitaev2006topological} proposed to use the tripartite information $I_3(A:B:C)\equiv S_A+S_B+S_C-S_{AB}-S_{AC}-S_{BC}+S_{ABC}\equiv I(A:B)+I(A:C)-I(A:BC)$, which is equal to $S_{\rm topo}$. If we study the same quantity for boundary regions $A,B,C$ in the RTN, it has contribution from not only the topological term but also geometrical terms. The correct quantity to study should be $I_3$ for three bulk contacting regions, but the physical interpretation at the boundary is not so clear to us at this moment. 

Another interesting comment is that bulk gravity is related to the boundary energy conservation, similar to how a bulk gauge symmetry is related to a boundary global symmetry. Therefore it is interesting to ask whether a similar approach can be used to construct tensor networks with certain energy conservation. The key difference between a global symmetry described by a Lie group and a boundary Hamiltonian is that the former acts in a tensor factorized way on the boundary. In the case of a Lie group, the conserved charge are Lie algebra generators of the form $T^a=\sum_{x\in B}T^a_x$. In contrast, a local boundary Hamiltonian is a sum over local terms which generically do not commute with each other. Is it possible to study tensor network states that preserve the energy of a given Hamiltonian? We will leave this for future work.

Ref. \cite{harlow2021symmetries} pointed out that when the boundary has a global symmetry, the bulk theory must contains the corresponding gauge theory coupled with matter fields that carry all irreducible representations of the global symmetry. In the RTN construction this is not required. The reason of this difference may tell us more about the difference between tensor networks and dynamical quantum gravity. 

\noindent{\bf Acknowledgement.} I would like to thank Xingshan Cui and Zhao Yang for discussions and collaboration on this project at the early stage. I would like to thank Michael Levin for helpful discussions. Some main results in this work have been reported in a lecture of mine in 2017.\cite{qi2017} I noticed an independent work\cite{morgan2021classical} that studied a similar symmetric random tensor network in the deconfined phase. This work is supported by the National Science Foundation under grant No. 2111998. This work is partially finished during my visit to the Institute for Advanced Study, Tsinghua University (IASTU). I would like to thank IASTU for hospitality.

\bibliographystyle{jhep}
\bibliography{refs}

\end{document}